# Buffer layer-less fabrication of high-mobility transparent oxide semiconductor, La-doped BaSnO₃

Anup V. Sanchela[1,*], Mian Wei[2], Joonhyuk Lee[3], Gowoon Kim[3], Hyoungjeen Jeen[3], Bin Feng[4], Yuichi Ikuhara[4], Hai Jun Cho[1,2], and Hiromichi Ohta[1,2*]

[1]Research Institute for Electronic Science, Hokkaido University, N20W10, Kita, Sapporo 001−0020, Japan

[2]Graduate School of Information Science and Technology, Hokkaido University, N14W9, Kita, Sapporo 060−0814, Japan

[3]Department of Physics, Pusan National University, Busan 46241, Korea

[4]Institute of Engineering Innovation, The University of Tokyo, 2−11−16 Yayoi, Bunkyo, Tokyo 113−8656, Japan

E-mail: anup.sanchela@es.hokudai.ac.jp, hiromichi.ohta@es.hokudai.ac.jp

**Transparent oxide semiconductors (TOSs) showing both high visible transparency and high electron mobility have attracted great attention towards the realization of advanced optoelectronic devices.[1-5] La-doped BaSnO₃ (LBSO) is one of the most promising TOSs because its single crystal exhibits a high electron mobility.[6-9] However, in the LBSO films, it is very hard to obtain high mobility due to the threading dislocations, which are originated from the lattice mismatch between the film and the substrate. Therefore, many researchers have tried to improve the mobility by inserting a buffer layer[6, 10-14]. While the buffer layers increased the electron mobilities, this approach leaves much to**

be desired since it involves a two-step film fabrication process and the enhanced mobility values are still significantly lower than single crystal values. We show herein that the electron mobility of LBSO films can be improved without inserting any buffer layers if the films are grown under highly oxidative ozone ($O_3$) atmospheres. The $O_3$ environments relaxed the LBSO lattice and reduced the formation of $Sn^{2+}$ states, which are known to suppress the electron mobility in LBSO.[15] The resultant $O_3$-LBSO films showed improved mobility values up to 115 cm² V⁻¹ s⁻¹, which is among the highest in LBSO films on $SrTiO_3$ substrates and comparable to LBSO films with buffer layers.

Perovskite transparent oxide $BaSnO_3$ has attracted great interest since substituting La in Ba sites turns it to an n-type degenerate semiconductor with a very high electrical conductivity and electron mobility. Studies have shown that La-doped $BaSnO_3$ ($La_xBa_{1-x}SnO_3$, LBSO) single crystals can exhibit a high electrical conductivity (~$10^4$ S cm⁻¹) and electron mobility (320 cm² V⁻¹ s⁻¹)[6, 7] with a carrier concentration of $8 \times 10^{19}$ cm⁻³ at room temperature (RT). Although the bandgap ($E_g$) of $BaSnO_3$ is ~3.1 eV, the $E_g$ of LBSO is ~3.5 eV due to the Burstein-Moss shift[8, 9]. Since these properties are attractive for designing next generation transparent thin film electronic devices,[1, 4, 5] the electrical transport properties of thin LBSO epitaxial films have been examined many times. However, the electron mobility values in LBSO films are widely scattered from 10 to 183 cm² V⁻¹ s⁻¹, which are much lower than the single crystal values.[6, 10-14]

Many reports attribute the mobility suppression to the charge carrier propagation hindrance at threading dislocations or grain boundaries, which are mainly caused by the lattice mismatch at the film/substrate interface[16-19]. Therefore, many researchers inserted thick insulating buffer layers between doped $BaSnO_3$ film and substrate to reduce the threading dislocations. For example, A. Prakash *et al.*[13] optimized the thickness of buffer and La doped BSO layer and found that the idea the buffer layer thickness was 124 nm; Shin *et al.*[20] used buffer layer thickness of 150 nm whereas Shiogai *et al.*[11] used 200 nm. Although these studies observed improved electron mobilities in LBSO films, the buffer layer insertion does not fully compensate the extra deposition step in the film fabrication process since the mobility values are still significantly lower compared to bulk single crystals. This can be an issue for practical device applications because one-step fabrication processes are much more preferred in mass productions. In our previous study[21], we observed that the electron mobility of epitaxial LBSO films was strongly suppressed for films thinner than 50 nm. The mobilities increased with increasing thickness and eventually saturated at 90−100 cm$^2$ V$^{-1}$ s$^{-1}$ (~200 nm on (001) MgO substrate and ~400 nm on (001) SrTiO$_3$ substrate).[21] This implies that the initial several hundred nanometers in LBSO films play as buffer layers. Furthermore, we detected 2+ valence states of Sn in the X-ray absorption spectroscopy (XAS) spectrum of an LBSO film, which should not be detected from stoichiometric LBSO and suggest the presence of oxygen vacancies.[15] Since the thickness dependence shows that the buffer region does not contribute much to the electron transport, the buffer region thickness needs to be reduced to achieve high mobility values without inserting buffer layers (**Fig. 1**).

In this study, we employ the results from our previous study[21] to design a one-step fabrication process for high-mobility LBSO films without inserting buffer layers. We assumed the main lattice defects in the mobility suppressed buffer region (several hundred nanometers) were oxygen vacancies, which can be reduced by depositing the film under highly oxidative atmosphere (**Fig. 1**). Therefore, we examined the relationship between the oxidation environment during the film growth and the electron transport properties of LBSO films. To reduce the formation of Sn$^{2+}$ ions and oxygen vacancies[15] in LBSO films, we injected ozone (O$_3$) during the film growth, which creates a much stronger oxidation environment compared to typically used oxygen gas (O$_2$).[22] The LBSO films grown in ozone (O$_3$-LBSO) indeed exhibited higher electron mobilities compared to the LBSO films deposited in oxygen ($O_2$-LBSO) without inserting any buffer layer, especially for films thinner than 150 nm. Despite being smaller than single crystal values, the best electron mobility observed in this study (115 $cm^2\ V^{-1}\ s^{-1}$) is among the highest in LBSO films on $SrTiO_3$ substrates and comparable to that of LBSO films deposited with buffer layers.[11, 13, 20]

2 %-La-doped LBSO epitaxial films were grown on (001) surface of $SrTiO_3$ substrates under $O_3$ and $O_2$ atmospheres using pulsed laser deposition (PLD, KrF excimer laser, $\lambda$= 248 nm, 10 Hz, Fluence ~2 J $cm^{-2}$ $pulse^{-1}$) technique. The substrate temperature was kept at 700 °C during the film growth, whereas the chamber pressures were 17 Pa for $O_3$-LBSO and 10 Pa for $O_2$-LBSO, respectively. In case of $O_2$-LBSO, only oxygen gas was injected in the chamber while the partial pressure of $O_3$ was set to 10% by an ozone generator for the synthesis of $O_3$-LBSO films. The films were annealed at 1200 °C in air to obtain the atomically smooth surfaces[11, 23], and the thicknesses of the films varied from 11 nm to 450 nm.

High-resolution X-ray diffraction measurements (XRD, Cu Kα₁, ATX-G, Rigaku Co.) were performed on the LBSO films to characterize their crystallographic properties in details. The film thicknesses were determined by analyzing Kiessing or Pendell esung fringes while the surface morphologies were observed using atomic force microscopy (AFM, Nanocute, Hitachi High-Tech Sci. Co.). All LBSO films were heteroepitaxially grown on the (001) $SrTiO_3$ substrates, which were confirmed by the XRD measurements. As shown in the **Supplementary Fig. S5**, stepped and terraced surfaces were observed in both $O_3$-LBSO and $O_2$-LBSO films.

In order to clarify the structural differences in the LBSO films grown under $O_3$ and $O_2$ atmospheres in detail, we measured the X-ray reciprocal space mappings (RSMs) of the films as shown in the **Supplementary Figs. S1-S4**. In both cases, an asymmetric 103 diffraction spot of BaSnO$_3$ was observed around the 103 diffraction spot of SrTiO$_3$. Using these RSMs, we extracted both in-plane (*a*-) and cross-plane (*c*-) lattice parameters as shown in **Fig. 2(a)**. Due to the large lattice mismatch between LBSO and SrTiO$_3$ (−5 %), the lattice parameter *a* was much smaller than the bulk value (0.4116 nm) whereas *c* was larger than the bulk value at the beginning of the film growth [**Fig .2(a)**]. The *a/c* increased/decreased with increasing the film thickness and they approached the bulk value (0.4116 nm). It should be noted that the O$_3$-LBSO showed larger differences from the bulk at the beginning of the film growth.

We then calculated the average lattice parameter $(a^2 \cdot c)^{1/3}$ as shown in **Fig. 2(b)**, which represents the bulk strain in the films. Although the $(a^2 \cdot c)^{1/3}$ of the O$_2$-LBSO films did not reach the bulk value, that of the O$_3$-LBSO increased dramatically with the thickness and almost reached the bulk value. These results suggest that the lattice relaxation of the LBSO films occurred quickly in the case of the O$_3$ atmosphere. However, the lateral grain size (*D*) of both atmosphere showed similar tendencies [**Fig. 2(c)**]; At the beginning of the film growth, the *D* of the LBSO films were ~30 nm, increased with thickness, and saturated when the thickness exceeds ~150 nm.

In order to further clarify the lattice strains, we observed the microstructure of the films using low angle annular dark-field scanning transmission electron microscopy (LAADF-STEM) as shown in **Fig. 3**. While columnar structures were seen in both the O$_3$-LBSO [**Fig. 3(a)**] and the O$_2$-LBSO films [**Fig. 3(b)**], the density of the columns in the O$_2$-LBSO film is higher. Since the contrast in LAADF image is sensitive to the strain field (i.e. bulk strain), these results suggest that lattice strain was be reduced under ozone atmosphere, which is in good agreement with the RSM results.

The Hall mobility ($\mu_{Hall}$), carrier concentration ($n$), and electrical conductivity ($\sigma$) of the LBSO films at RT were measured using the conventional dc four-probe method with van der Pauw electrode configuration. Thermopower ($S$) was measured at RT from the thermo-electromotive force ($\Delta V$) generated by a temperature difference $\Delta T$ of ~4 K across the film, which was created using two Peltier devices. The temperatures at each end of the films were simultaneously measured with two thermocouples, and the $S$-values were calculated from the slope of the $\Delta V$–$\Delta T$ plots (correlation coefficient: >0.9999).

The RT electron transport properties of the $O_3$-LBSO and $O_2$-LBSO films against the film thicknesses are summarized in **Fig. 4**. The charge carrier concentration ($n$) in the LBSO films increased with the thickness and approached the nominal $n_{nom}$, which is defined by the atomic concentration of the La-dopants ($\equiv$[2%-La$^{3+}$] = 2.87 × 10$^{20}$ cm$^{-3}$). It should be noted that the $n$ of $O_3$-LBSO films exhibited a much faster increase compared to $O_2$-LBSO films. For example, the $n$ of ~100-nm-thick $O_3$-LBSO film is 87 % of the $n_{nom}$ while that $n$ of the $O_2$-LBSO film with the same thickness was only 46 % of the $n_{nom}$ **[Fig. 4(a)]**.

Similar trends were observed in the Hall mobility ($\mu_{Hall}$) and the electrical conductivity ($\sigma$) of the LBSO films (i.e. faster increase with thickness in $O_3$-LBSO); the $\mu_{Hall}$ of ~100-nm-thick $O_3$-LBSO film was 103 cm$^2$ V$^{-1}$ s$^{-1}$, which is almost two times higher than 57 cm$^2$ V$^{-1}$ s$^{-1}$ for the $O_2$-LBSO film with the same thickness. The effect of the $O_3$ atmosphere can be clearly seen as $O_3$-LBSO films always exhibited significantly higher values compared to $O_2$-LBSO films at the same thickness **[Figs. 4(b) and 4(c)]**. The highest $\mu_{Hall}$ value (115 cm$^2$ V$^{-1}$ s$^{-1}$) was observed in the 452-nm-thick $O_3$-LBSO film. In addition, the thermopower ($-S$) of the LBSO films decreased with increasing thickness **[Fig. 4(d)]**, and the $S$ declination rate of $O_3$-LBSO films was much higher than that of $O_2$-LBSO films **[Fig. 4(d)]**. Since |$S$| decreases with

$n$,[24] the observed $-S$ reduction is consistent with the thickness dependence of $n$ [**Fig. 4(a)**]. From these results, we conclude that LBSO films grown under $O_3$ atmosphere showed improved electron transport properties.

To validate our initial hypothesis on the role of oxygen deficiencies on the formation of $Sn^{2+}$ states,[15] we performed X-ray absorption spectroscopy (XAS) around the Sn $M_{4,5}$ edge of ~100-nm-thick LBSO films on $SrTiO_3$ substrate in Pohang accelerator laboratory (2A). We used surface sensitive electron yield mode (the penetration depth of X-ray ~10 nm). The XAS spectra of the LBSO films are shown in **Fig. 5**. In both films, four absorption peaks from stoichiometric $Sn^{4+}$ ions were observed [**Fig. 5(a)**], which are in good agreements with other studies.[25-27] Interestingly, just as we intended, $Sn^{2+}$ peak at ~488 eV[25, 28, 29] dissapeared in the $O_3$-LBSO film and can only be seen from the XAS of $O_2$-LBSO film [**Fig. 5(a)**]. The thicknesses of the $O_2$-LBSO and $O_3$-LBSO films were 107 nm and 125 nm, respectively. Despite the similar thickness values, the $\mu_{Hall}$ of the $O_2$-LBSO film (57 $cm^2$ $V^{-1}$ $s^{-1}$) was much lower compared to that of the $O_3$-LBSO film (103 $cm^2$ $V^{-1}$ $s^{-1}$). This result suggests that the existence of the $Sn^{2+}$ ion can indeed be related to the mobility suppression of LBSO films.[15]

In the stoichiometric composition of $BaSnO_3$, the stable valence state of Sn is 4+. If the two extra electrons from an oxygen vacancy form $Sn^{2+}$ ion, the $Sn^{2+}$ ions may degrade the carrier generation efficiency of $La^{3+}$ ions and scatter electrons. Therefore, it is plausible for oxygen deficiencies to reduce the transport properties of electrons in LBSO. In fact, MBE-grown LBSO films with high mobility were fabricated using highly oxidative atmosphere.[12, 13, 14] Thus, in the context of our current study, it can be suspected that low mobility and low carrier concentration in LBSO films observed in other studies are related to the non-oxidative atmosphere. The mobility dependence on the thickness suggests that these defects are concentrated at the film/substrate interface.[21] However, we would like to note that the formation of $Sn^{2+}$ states and oxygen vacancies in LBSO needs to be studied much more thoroughly to clarify their exact location in the film.

In summary, we have successfully improved the electron mobility of La-doped $BaSnO_3$ films without inerting buffer layers by using highly oxidative atmosphere ($O_3$) during the film growth. The lattice relaxation of the $O_3$-LBSO films occurred when the film thickness was thinner than the $O_2$-LBSO films. Despite being smaller than single crystal values, the highest electron mobility observed in this study (115 $cm^2$ $V^{-1}$ $s^{-1}$) is among the highest in LBSO films on $SrTiO_3$ substrates and similar to LBSO films with buffer layers[11, 13, 20]. The XAS results showed that the $Sn^{2+}$ concentration in the LBSO films was successfully reduced by creating an oxidative atmosphere during the film growth with the injection of $O_3$.

**Acknowledgements**

This research was supported by Grants-in-Aid for Scientific Research on Innovative Areas "Nano Informatics" (25106003 and 25106007) from the Japan Society for the Promotion of Science (JSPS). H.J. and H.O. are supported by the Korea-Japan bilateral program funded from following programs of each country: International cooperation program by the NRF (NRF-2018K2A9A2A08000079) and JSPS. H.O. is supported by Grants-in-Aid for Scientific Research A (17H01314) from the JSPS, the Asahi Glass Foundation, and the Mitsubishi Foundation. A part of this work was supported by Dynamic Alliance for Open Innovation Bridging Human, Environment and Materials, and by the Network Joint Research Center for Materials and Devices.

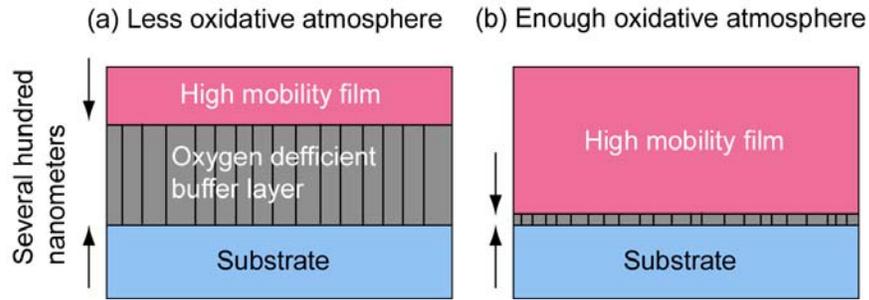

**FIG. 1 | Hypothesis: Schematic illustration of La:BaSnO₃ films.** (a) La:BaSnO$_3$ film grown under less oxidative atmosphere. Since the buffer layer is not fully oxidized, several hundred nanometer thick oxygen deficient buffer layer is initially grown on the substrate. After the lattice relaxation, high mobility La:BaSnO$_3$ film growth occurs. (b) La:BaSnO$_3$ film grown under enough oxidative atmosphere. A fully oxidized buffer layer is initially formed on the substrate. The lattice relaxation occurs in the thinner film, after that, high mobility La:BaSnO$_3$ film growth occurs. We hypothesize the buffer layer thickness should be very high to achieve the high mobility for low oxidative atmosphere. Because the buffer contains oxygen deficiency (a). On the other hand, buffer layer thickness highly reduced by using enough oxidative atmosphere to acquiring high mobility (b), which will be useful for practical application.

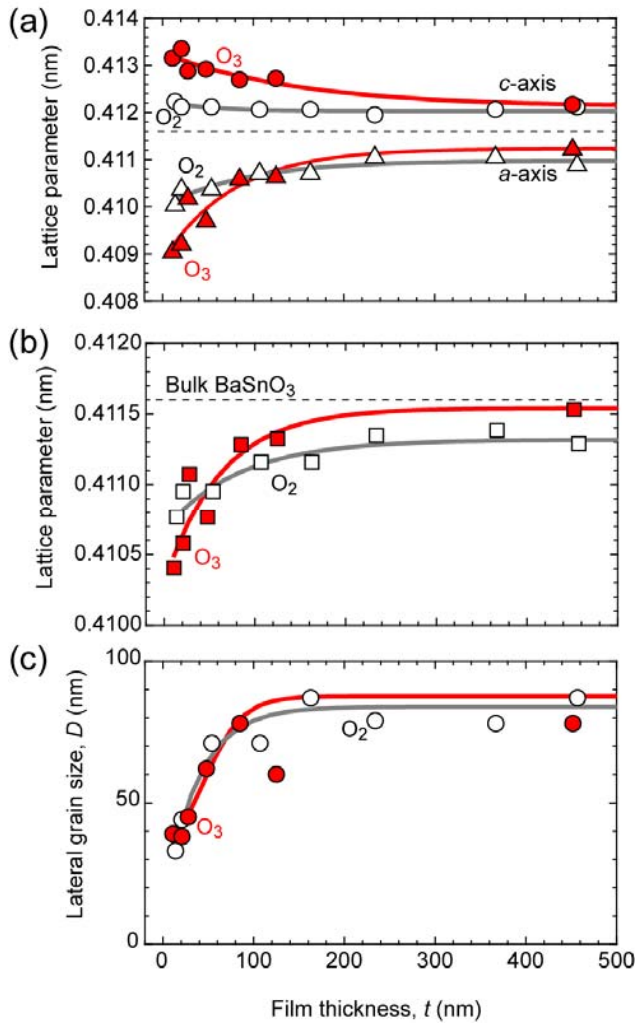

**FIG. 2 | Lattice characteristics of the La:BaSnO₃ films at RT.** Thickness dependences of (a) lattice parameters, *a* and *c*, (b) average lattice parameter $(a^2 \cdot c)^{1/3}$, and lateral grain size for the resultant LBSO films grown under ozone (filled symbols) and oxygen (open symbols, Ref. 21) atmospheres. Average lattice parameter of the O₃-LBSO increased dramatically with the thickness and almost reached the bulk value (0.4116 nm).

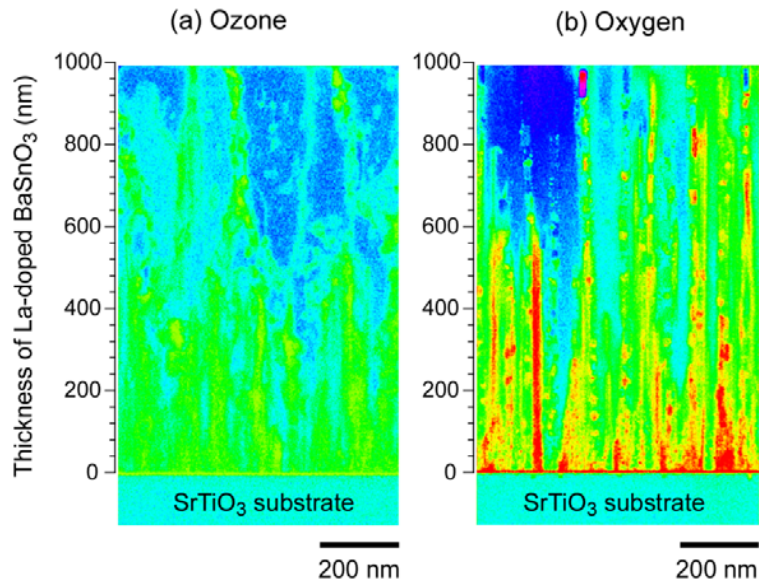

**FIG. 3 | Cross-sectional LAADF-STEM images of the LBSO films.** Columner structure is seen in both (a) the O$_3$-LBSO and (b) the O$_2$-LBSO films (Ref. 21)[21]. The contrast in LAADF image is sensitive to the strain field. The density of the strain field in the O$_2$-LBSO film is higher.

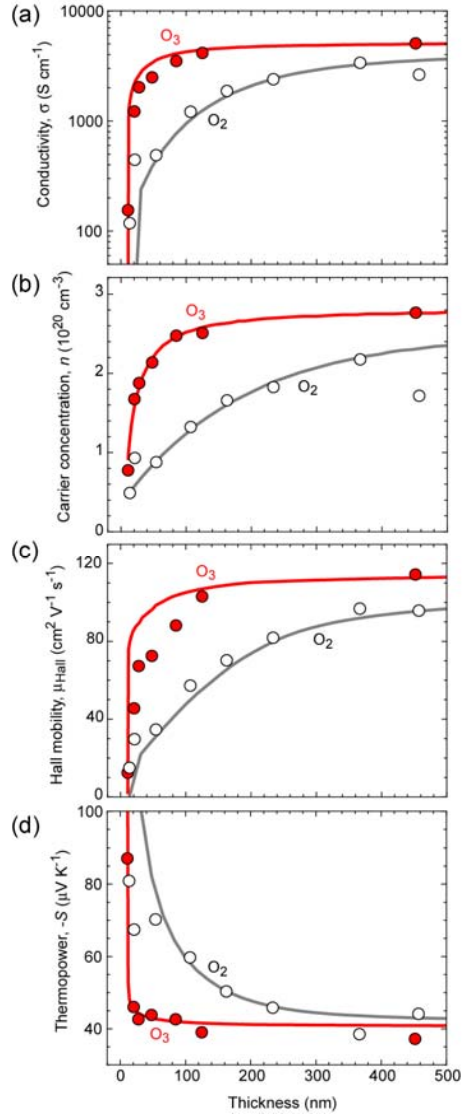

**FIG. 4 | Electron transport properties of the resultant La:BaSnO₃ films at room temperature.** (a) electrical conductivity, $\sigma$, (b) carrier concentration, $n$, (c) Hall mobility, $\mu_{Hall}$, and (d) thermopower, $S$ of the LBSO films grown under ozone atmosphere (filled circle) and oxygen atmosphere (open circle, Ref. 21). The thickness dependences of the $n$ were fitted by using the Boltzmann function as shown by the lines. The $\mu_{Hall}$ reaches 115 cm² V⁻¹ s⁻¹ under ozone atmosphere. The $n$, $\sigma$, and $\mu_{Hall}$, of the LBSO films grown under the ozone atmosphere show a drastic increase and saturation in the thin (< 150 nm) region compared with those grown under the oxygen atmosphere.

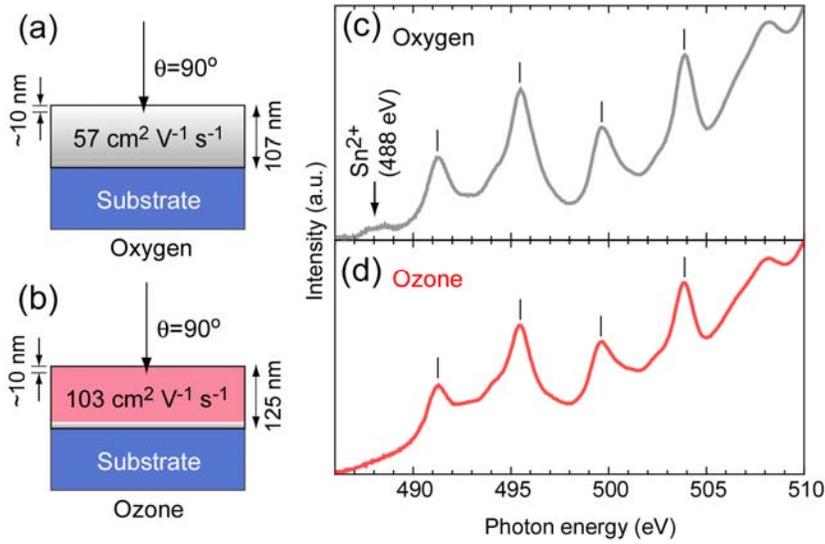

**FIG. 5 | XAS spectra of the LBSO films.** Schematic cross-sectional structure and XAS spectra at around the Sn $M_{4,5}$ edge of the LBSO films. (a)(c) 107-nm-thick LBSO film grown under the oxygen atmosphere ($\mu_{\text{Hall}}$=57 cm$^2$ V$^{-1}$ s$^{-1}$). The penetration depth of the X-ray is ~10 nm. A broad absorption peak of Sn$^{2+}$ is clearly seen at 488 eV. (b)(d) 125-nm-thick LBSO film grown under the ozone atmosphere ($\mu_{\text{Hall}}$=103 cm$^2$ V$^{-1}$ s$^{-1}$). Only intense absorption peaks of Sn$^{4+}$ are seen as shown using bars. Note: the Sn $M$ edge is split into $3d_{3/2}$ ($M_4$-edge) and $3d_{5/2}$ ($M_5$-edge) due to electronic dipole transitions. The two peaks located between 491 eV and 496 eV are the peaks corresponding to $M_5$-edge, while the other two peaks located between 500 eV and 505 eV are the peaks from $M_4$-edge.